\documentclass[showpacs,preprintnumbers,prd,nofootinbib,floats,amssymb,floatfix]{revtex4}
\usepackage{amsmath}
\usepackage{graphicx}
\usepackage{amsxtra}
\usepackage{hyperref}
\usepackage{amsmath}
\usepackage{amstext}
\begin{document}
\title{ The Hamilton-Jacobi characteristic equations  for topological invariants: Pontryagin and Euler classes  }
\author{Alberto Escalante}  \email{aescalan@ifuap.buap.mx}
\author{ Aldair-Pantoja}  \email{jpantoja@ifuap.buap.mx}
 \affiliation{  Instituto de F{\'i}sica, Benem\'erita Universidad Aut\'onoma de Puebla. \\
 Apartado Postal J-48 72570, Puebla Pue., M\'exico, }
\begin{abstract}
By using the Hamilton-Jacobi [HJ]  framework    the topological theories  associated with   Euler and Pontryagin classes are analyzed. We report the construction of a fundamental  $HJ$ differential where the characteristic equations and the  symmetries of the theory are identified. Moreover, we work in both theories with the same phase space variables and we  show that in spite of   Pontryagin  and  Euler classes share the same equations of motion   their   symmetries are different.   In addition, we report all  HJ Hamiltonians and we  compare our results with other formulations  reported in the literature.   
\end{abstract}
 \date{\today}
\preprint{}
\maketitle
\section{Introduction}
It is well-known that the study of topological theories is  an interesting topic  in mathematics or physics from either or both directions. In fact,  for the former the study of topological structures   of manifolds (with particular interest in four dimensional manifolds) based on the identification of topological invariants is one of the research subjects for the mathematical physics community.  Examples   of topological invariants in four dimensions can be cited,    the so-called Euler  and  Pontryagin invariants. These invariants are
fundamental objects in the characterization of the topological structure of a manifold, they label topologically distinct four-geometries; the Pontryagin invariant gives the relation between the number of selfdual and anti-selfdual harmonic connections on the manifold. Moreover, the Euler invariant   gives a relation between the number of harmonic p-forms on the manifold \cite{1}. From the physical point of view, there exist a close relation between these topological invariants and physical theories just like  gravity  and field theory. In this respect,    the Euler  and  Pontryagin classes   are   fundamental blocks for constructing the noncommutative form of topological gravity  \cite{2}. In the $BF$-gravity context  there exists also a   relation  between  these invariants and $BF$-gravity formulation. In fact, in the MacDowell-Mansouri formulation of gravity based on a $SO(5)$ topological $BF$ theory,   the symmetry group  SO(5)  is broken  down into  SO(4) in order to obtain  the Palatini action plus the addition  of the Pontryagin  and Euler topological invariants \cite{3, 4, 5}, due to  these topological classes have trivial local variations that do not contribute  to the dynamics, hence one obtains essentially general relativity. In this respect,  both the   Euler  and   Pontryagin invariants can be written as a $BF$-like theory  and this fact has   allowed   to study its canonical structure in more convenient way \cite{6}. In addition, the study of the canonical structure of Pontryagin invariant   in the presence of boundaries has been analyzed in   \cite{8}. In fact,  topological gauge theories defined   on spacetime regions with boundary  are good toy models  for studying the emerging  of physical degrees of freedom and  observables localized on the boundary; these theories are  natural objects for researching  the   relation  between the canonical structure of topological gauge theory and  the existence of dualities a la $AdS/CFT$ \cite{9, 10, 11}. \\
On the other hand, from the Hamiltonian point of view, the Euler  and  Pontryagin invariants  treated   as field theories give rise   the same equations of motion, are devoid of physical degrees of freedom,  diffeomorphisms covariant and there exist reducibility conditions between the constraints \cite{7a}. Furthermore, form the quantum point of view due to  the  $BF$ structure of these topological theories  and the  symmetries commented above,   these  invariants are good laboratories for studying the classical and quantum structure of a  background independent theory  and this fact could contribute to  the spin foam formulation developed for $BF$ theories in the Loop Quantum Gravity [LQG] program \cite{12}.\\
With all these ideas in mind, the purpose of this paper is to develop the Hamilton-Jacobi  [HJ] analysis of  the Euler and Pontryagin invariants. As far as we know the Hamilton-Jacobi analysis  of these invariants has not been carried  out. In this respect,  we wish to extend the results reported in \cite{7a, 13} where these invariants within  the Dirac and Fadeev-Jackiw context have been analyzed. We shall  use the Hamilton-Jacobi [HJ] scheme developed by G\"uler  \cite{F17, F18, F19, F20, F21a} because is a  good alternative  for analyzing  gauge systems. In fact, the G\"uler approach is based on the construction of a fundamental differential defined on the full  phase space, and the elementary  blocks for constructing the fundamental differential are the constraints of the theory called Hamiltonians. The HJ Hamiltonians can be involutives or noninvolutives and they are elementary  for obtaining the characteristic equations from which  one can identify  the gauge symmetries,  the equations of motion and the physical degrees of freedom. The construction of the fundamental differential is direct; in general the Hamiltonians of HJ approach do not coincide with the constraints obtained in the Dirac formalism,  and   the process for identifying the symmetries in HJ framework is  more economical than other approaches; in this sense the HJ framework is a good  alternative for analyzing gauge systems.\\
The paper is organized as follows.  In Section II, the HJ analysis  of  the Pontryagin invariant is performed. We will write the invariant in a  $BF$-like form, which    will be useful in order to introduce a new set of variables that will allow us to compare the symmetries of both invariants  in a better form. We identify all Hamiltonians of the theory and the  fundamental differential is constructed, then all symmetries of the theory are identified. In Section III, we  rewrite the Euler invariant also in a $BF$-like form  and we use the same variables  introduced in the Section I. Then the HJ analysis is performed; we construct the fundamental differential of the theory and the characteristic equations are obtained. We compare the symmetries of  both  theories and  we will show that in spite of the invariants share the same equations of motion  their  symmetries  are different  to each  other. Finally, in Section IV we add some remarks and conclusions.
\section{Hamilton-Jacobi analysis of the Pontryagin invariant}
It is well-known that the Pontryagin invariant is given by the following action 
\begin{equation}
S[A_{\mu}{}^{IJ}]=\Xi \int_{M} F^{IJ}\wedge F_{IJ}, 
\label{1}
\end{equation}
where $\Xi$ is a constant, $M$ is a four-dimensional manifold without boundary, $I, J, K=0, 1, 2, 3$ are $SO(3,1)$ indices that can be raised and lowered  with the  metric $\eta_{IJ}=(-1,1, 1,1)$, $F_{\alpha\beta}{}^{IJ}$ is the strength curvature  of the 1- form connexion $A^{IJ}$  defined as $F_{\alpha\beta}{}^{IJ}=\partial_{\alpha}A_{\beta}^{IJ}-\partial_{\beta}A_{\alpha}{}^{IJ}+A_{\alpha}{}^{I}{}_{K}A_{\beta}{}^{KJ}-A_{\alpha}{}^{J}{}_{K}A_{\beta}{}^{KI}$.  The action  (\ref{1}) was analyzed in \cite{7a, 13} by using the Dirac and Faddeev-Jackiw  approaches respectively,  in those papers were reported that the system presents first and second class constraints and a symplectic tensor was constructed, however, the symmetries  of the constraints were not reported. Moreover, a direct comparison between Pontryagin and Euler invariants was not performed. We can observe that in the action (\ref{1})  the dynamical variable is  the connection $A$ and $F$ is a label; we can write the action (\ref{1}) in a new fashion form, in a $BF$-like theory, this step is convenient for future computations,  for which   a new  set of variables  will be introduced  allowing  us to perform the comparison of both theories under study   in  more convenient form. The Pontryagin invariant  in a $BF$ fashion  is expressed  by 
\begin{equation}
S[A_{\mu}{}^{IJ},B_{\mu\nu}{}^{IJ}]=\Xi\int_{M}[F^{IJ}\wedge B_{IJ}-\frac{1}{2}
B^{IJ}\wedge B_{IJ}], 
\label{2a}
\end{equation}
here $B^{IJ}=\frac{1}{2}B_{\alpha\beta}{}^{IJ}dx^{\alpha}\wedge dx^{\beta}$ is a set of six two-forms.  From (\ref{2a}) the following equations of motion arise 
\begin{eqnarray}
DB&=&0, \nonumber\\
F&=&B, 
\label{3}
\end{eqnarray}
by taking  into account  (\ref{3}) in (\ref{2a}) we obtain again (\ref{1}), from this point of view either $A$ or $B$ are dynamical fields and this fact will be taken into account in the analysis. 
By performing the  3+1 decomposition and breaking down the Lorentz covariance   we obtain the following Lagrangian
\begin{eqnarray}
\nonumber \mathcal{L}&=&\Xi\eta^{abc}\int_{M}\Big[
B_{bc0i}\dot{A}_{a}{}^{0i}+\frac{1}{2}B_{bcij}\dot{A}_{a}{}^{ij}
+\frac{1}{2}\left(\partial_{a}B_{bcij}+2B_{bcik}A_{aj}{}^{k}+2B_{bcoi}A_{a}{}^{0}{}_{j}\right)A_{0}{}^{ij} \\ \nonumber 
&+&(\partial_{a}B_{bc0i}+B_{bcij}A_{a0}^{j}+B_{bc0j}A_{a}{}^{j}{}_{i})A_{0}{}^{0i}
+(\partial_{a}A_{b}{}^{0i}-\partial_{b}A_{a}{}^{0i}+A_{a}{}^{i}{}_{j}A_{b}{}^{0j}-A_{b}{}^{i}{}_{j}A_{a}{}^{0j})B_{0c0i} \\ \nonumber 
&+& \frac{1}{2}(\partial_{a}A_{b}{}^{ij}-\partial_{b}A_{a}{}^{ij}
+A_{a}{}^{i}{}_{0}A_{b}{}^{0j}-A_{b}{}^{i}{}_{0}A_{a}{}^{0j}
+A_{a}{}^{i}{}_{k}A_{b}{}^{kj}-A_{b}{}^{i}{}_{k}A_{a}{}^{kj})B_{0cij} \\
&-&\frac{1}{4}(B_{0a}{}^{ij}B_{bcij}+B_{abij}B_{0cij})
-\frac{1}{2}(B_{0a}{}^{0i}B_{bc0i}+B_{ab}{}^{0i}B_{0c0i}) \Big]d^{3}x, 
\end{eqnarray}
now we introduce   the following variables \cite{13}
\begin{eqnarray}
\nonumber A_{aij}&\equiv&-\epsilon_{ijk}A_{a}{}^{k}, \\ \nonumber 
A_{0ij}&\equiv&-\epsilon_{ijk}A_{0}{}^{k}, \\ \nonumber 
B_{abij}&\equiv&-\epsilon_{ijk}B_{ab}{}^{k}, \\ \nonumber 
B_{0aij}&\equiv&-\epsilon_{ijk}B_{0a}{}^{k}, \\  \nonumber 
A_{ai}&\equiv&\Upsilon_{ai},  \\ \nonumber
A_{0}{}^{i}&\equiv& -T^{i}, \\ \nonumber
A_{00i}&\equiv& -\Lambda_{i}, \\ \nonumber
B_{0a}{}^{0i}&\equiv& - \frac{1}{2}\varsigma_{a}{}^{i}, \\  \nonumber 
B_{0ai}&\equiv& -\frac{1}{2}\chi_{ai},
\end{eqnarray}
and the Lagrangian is rewritten in the following form  
\begin{eqnarray}
\nonumber
\mathcal{L}&=&\Xi\eta^{abc}\int_{M}\Big[
B_{ab0i}\dot A_{c}{}^{0i}+B_{abi}\dot{\Upsilon}_{c}{}^{i}
-(\partial_{c}B_{abk}-\epsilon^{ij}{}_{k}B_{abi}\Upsilon_{cj}-\epsilon^{ij}{}_{k}B_{ab0i}A_{c}{}^{0}{}_{j})T^{k} 
\\
\nonumber
&-&(\partial_{c}B_{ab0i}-\epsilon_{ij}{}^{k}B_{abk}A_{c0}{}^{j}+\epsilon_{i}{}^{j}{}_{k}B_{ab0j}\Upsilon_{c}{}^{k})\Lambda^{i} 
\\
\nonumber
&-&\frac{1}{2}(\partial_{b}A_{c0}{}^{i}-\partial_{c}A_{b0}{}^{i}+\epsilon^{i}{}_{jk}A_{b0}{}^{j}\Upsilon_{c}{}^{k}-\epsilon^{i}{}_{jk}A_{c0}{}^{j}\Upsilon_{b}{}^{k}+B_{bc}{}^{0i})\varsigma_{ai} 
\\
&-&\frac{1}{2}(\partial_{b}\Upsilon_{ci}-\partial_{c}\Upsilon_{bi}-\epsilon_{ijk}A_{b0}{}^{j}A_{c0}{}^{k}+\epsilon_{ijk}\Upsilon_{b}{}^{j}\Upsilon_{c}{}^{k}-B_{abi})\chi_{a}{}^{i}\Big]d^{3}x.
\end{eqnarray}
From the definition of the momenta  $(p^{a0i}, \pi^{ai}, \hat{T}^{i}, \hat{\Lambda}^{i}, \hat{\varsigma}^{ai}, \hat{\chi}^{ai}, p^{ab0i}, p^{abi} )$ canonically conjugate to $({A}_{a0i}, {\Upsilon}_{ai}, {T}_{i}, {\Lambda}_{i}, {\varsigma}_{i}, {\chi}_{ai}, {B}_{ab0i}, {B}_{abi}  )$ 
we identify the following $HJ$ Hamiltonians of the theory
\begin{eqnarray}
\nonumber
H'&\equiv&\Pi+H_{0}=0, 
\\
\nonumber
\phi_{1}^{a0i}&\equiv& p^{a0i}-\Xi\eta^{abc}B_{bc}{}^{0i}=0,
\\
\nonumber
\phi_{2}^{ai}&\equiv& \pi^{ai}-\Xi\eta^{abc}B_{bc}{}^{i}=0,
\\
\nonumber
\phi_{3}^{i}&\equiv&\hat{T}^{i}=0,
\\
\nonumber
\phi_{4}^{i}&\equiv&\hat{\Lambda}^{i}=0,
\\
\nonumber
\phi_{5}^{ai}&\equiv&\hat{\varsigma}^{ai}=0,
\\
\nonumber
\phi_{6}^{ai}&\equiv&\hat{\chi}^{ai}=0,
\\
\nonumber
\phi_{7}^{ab0i}&\equiv& p^{ab0i}=0,
\\
\phi_{8}^{abi}&\equiv& p^{abi}=0,
\end{eqnarray}
where $\Pi=\partial_{0}S$,   $S$ is the action  and $H_{0}$ is identified as the canonical Hamiltonian expressed by 
\begin{eqnarray}
\nonumber
H_{0}&=&(\partial_{a}\pi_{a}{}^{i}-\epsilon^{i}{}_{jk}\pi^{aj}\Upsilon_{a}{}^{k}-\epsilon^{i}{}_{jk}p^{a0j}A_{a0}{}^{k})T_{i}
+(\partial_{a}p^{a0i}+\epsilon^{i}{}_{jk}\pi^{aj}A_{a0}{}^{k}-\epsilon^{i}{}_{jk}p^{a0j}\Upsilon_{a}{}^{k})\Lambda_{i}
\\
\nonumber
&+&\frac{\Xi}{2}\eta^{abc}(\partial_{b}A_{c0}{}^{i}-\partial_{c}A_{b0}{}^{i}+\epsilon^{i}{}_{jk}A_{b0}{}^{j}\Upsilon_{c}{}^{k}-\epsilon^{i}{}_{jk}A_{c0}{}^{j}\Upsilon_{b}{}^{k})\varsigma_{ai}
\\
\nonumber
&+&\frac{\Xi}{2}\eta^{abc}(\partial_{b}\Upsilon_{c}{}^{i}-\partial_{c}\Upsilon_{b}{}^{i}-\epsilon^{i}{}_{jk}A_{b0}{}^{j}A_{c0}{}^{k}+\epsilon^{i}{}_{jk}\Upsilon_{b}{}^{j}\Upsilon_{c}{}^{k})\chi_{ai}
\\
&+&\frac{1}{2}\varsigma_{ai}p^{a0i}-\frac{1}{2}\chi_{a}{}^{i}\pi_{ai},
\end{eqnarray}
now, from the definition of the momenta  we also identify the  fundamental Poisson brackets between dynamical variables 
\begin{eqnarray}
\nonumber
\lbrace A_{a0i}(x),p^{b0j}(y)\rbrace &=&\delta_{a}^{b}\delta_{i}^{j}\delta^{3}(x-y),
\\
\nonumber
\lbrace \Upsilon_{ai}(x),\pi^{bj}(y)\rbrace &=&\delta_{a}^{b}\delta_{i}^{j}\delta^{3}(x-y),
\\
\nonumber
\lbrace T_{i}(x),\hat{T}^{j}(y)\rbrace &=&\delta_{a}^{b}\delta_{i}^{j}\delta^{3}(x-y),
\\
\nonumber
\lbrace \Lambda_{i}(x),\hat{\Lambda}^{j}(y)\rbrace &=&\delta_{a}^{b}\delta_{i}^{j}\delta^{3}(x-y),
\\
\nonumber
\lbrace \varsigma_{ai}(x),\hat{\varsigma}^{bj}(y)\rbrace &=&\delta_{a}^{b}\delta_{i}^{j}\delta^{3}(x-y),
\\
\nonumber
\lbrace \chi_{ai}(x),\hat{\chi}^{bj}(y)\rbrace &=&\delta_{a}^{b}\delta_{i}^{j}\delta^{3}(x-y),
\\
\nonumber
\lbrace B_{ab0i}(x),p^{cd0j}(y)\rbrace&=&\frac{1}{2}(\delta_{a}^{c}\delta_{b}^{d}-\delta_{a}^{d}\delta_{b}^{c})\delta_{i}^{j}\delta^{3}(x-y),
\\
\lbrace B_{abi}(x),p^{cdj}(y)\rbrace &=&\frac{1}{2}(\delta_{a}^{c}\delta_{b}^{d}-\delta_{a}^{d}\delta_{b}^{c})\delta_{i}^{j}\delta^{3}(x-y).
\end{eqnarray}
In this manner, with the  Hamiltonians at hand, we construct the fundamental differential which describes the evolution of any  function, say $f$, on the phase space \cite{ F17, F18, F19, F20}
\begin{eqnarray}
\nonumber
df(x)&=&\int d^{3}y\Big[\lbrace f(x),H'(y)\rbrace dt
+\lbrace f(x),\phi_{1}^{a0i} \rbrace d\rho_{a0i}
+\lbrace f(x),\phi_{2}^{ai} \rbrace d\tilde {\varphi}_{ai}
+\lbrace f(x),\phi_{3}^{i} \rbrace d\tau_{i}
+\lbrace f(x),\phi_{4}^{i} \rbrace d\lambda_{i}
\\
&+&\lbrace f(x),\phi_{5}^{ai} \rbrace d\sigma_{ai}
+\lbrace f(x),\phi_{6}^{ai} \rbrace d\zeta_{ai}
+\lbrace f(x),\phi_{7}^{ab0i} \rbrace d\theta_{ab0i}
+\lbrace f(x),\phi_{8}^{abi} \rbrace d\xi_{abi}
\Big],
\end{eqnarray}
where $(\rho_{a0i}, \tilde {\varphi}_{ai}, \tau_{i}, \lambda_{i}, \sigma_{ai}, \zeta_{ai}, \theta_{ab0i}, \xi_{abi})$ are parameters associated with the Hamiltonians.  On the other hand, we observe that the Hamiltonians $\phi_{3}^{i}$, $\phi_{4}^{i}$, $\phi_{5}^{ai}$ and $\phi_{6}^{ai}$  are involutives and $\phi_{1}^{a0i}$, $\phi_{2}^{ai}$ $\phi_{7}^{ab0i}$ and $\phi_{8}^{abi}$  are non-involutives. Involutive Hamiltonians, are those whose Poisson brackets with all Hamiltonians,  including themselves,      vanish; otherwise, they are called non-involutives. The presence of non-involutive Hamiltonians introduce the generalized $HJ$ brackets defined by \cite{ F17, F18, F19, F20}
\begin{align}
\{A, B\}^{*} &=\{A, B\} - \{A, H'_{\bar{a}}\}(C{_{\bar{a}\bar{b}}})^{-1}\{H'_{\bar{b}}, B\},
\label{10}
\end{align}
where $(C{_{\bar{a}\bar{b}}})$ is the matrix whose entries are given by  the Poisson brackets between non-involutives Hamiltonians and $(C{_{\bar{a}\bar{b}}})^{-1}$ its inverse matrix; explicitly 
\begin{equation}
C_{\bar{a}\bar{b}}
=
\begin{pmatrix}
0 & 0 & \Xi\eta^{abc}\delta^{il} & 0 \\
0 & 0 & 0 & -\Xi\eta^{abc}\delta^{il} \\
-\Xi\eta^{abc}\delta^{il} & 0 & 0 & 0 \\
0 & \Xi\eta^{abc}\delta^{il} & 0 & 0 \\
\end{pmatrix}
\delta^{3}(x-y),
\end{equation}

and 

\begin{equation}
C_{\bar{a}\bar{b}}^{-1}
=
\begin{pmatrix}
0 & 0 & -\frac{1}{2\Xi}\eta^{def}\delta_{lj} & 0 \\
0 & 0 & 0 & \frac{1}{2\Xi}\eta^{def}\delta_{lj} \\
\frac{1}{2\Xi}\eta_{def}\delta_{lj} & 0 & 0 & 0 \\
0 & -\frac{1}{2\Xi}\eta_{def}\delta_{lj} & 0 & 0 \\
\end{pmatrix}
\delta^{3}(x-y),
\end{equation}
hence, the generalized brackets between the dynamical variables read 
\begin{eqnarray}
\nonumber
\lbrace A_{a0i}(x),p^{b0j}(y)\rbrace ^{*} &=&\delta_{a}^{b}\delta_{i}^{j}\delta^{3}(x-y),
\\
\nonumber
\lbrace \Upsilon_{ai}(x),\pi^{bj}(y)\rbrace ^{*} &=&\delta_{a}^{b}\delta_{i}^{j}\delta^{3}(x-y),
\\
\nonumber
\lbrace T_{i}(x),\hat{T}^{j}(y)\rbrace ^{*} &=&\delta_{a}^{b}\delta_{i}^{j}\delta^{3}(x-y),
\\
\nonumber
\lbrace \Lambda_{i}(x),\hat{\Lambda}^{j}(y)\rbrace ^{*} &=&\delta_{a}^{b}\delta_{i}^{j}\delta^{3}(x-y),
\\
\nonumber
\lbrace \varsigma_{ai}(x),\hat{\varsigma}^{bj}(y)\rbrace ^{*} &=&\delta_{a}^{b}\delta_{i}^{j}\delta^{3}(x-y),
\\
\nonumber
\lbrace \chi_{ai}(x),\hat{\chi}^{bj}(y)\rbrace ^{*} &=&\delta_{a}^{b}\delta_{i}^{j}\delta^{3}(x-y),
\\
\nonumber
\lbrace B_{ab0i}(x),p^{cd0j}(y)\rbrace ^{*} &=&0,
\\
\nonumber
\lbrace B_{abi}(x),p^{cdj}(y)\rbrace ^{*} &=&0,
\\
\nonumber
\lbrace B_{ab0i}(x),A_{c0j}(y)\rbrace ^{*} &=&\frac{1}{2\Xi}\eta_{abc}\delta_{ij}\delta^{3}(x-y),
\\
\lbrace B_{abi}(x),\Upsilon_{cj}(y)\rbrace ^{*} &=&-\frac{1}{2\Xi}\eta_{abc}\delta_{ij}\delta^{3}(x-y),
\end{eqnarray}
where we observe that there is a contribution in these brackets due to  the parameter $\Xi$. The introduction of the generalized brackets allows  us to  rewrite the fundamental differential in terms of involutives Hamiltonians \cite{F17, F18, F19, F20, F21a}
\begin{eqnarray}
\nonumber
df(x)&=&\int d^{3}y\Big[\lbrace f(x),H'(y)\rbrace ^{*} dt
+\lbrace f(x),\phi_{3}^{i}(y) \rbrace ^{*} d\tau_{i}
+\lbrace f(x),\phi_{4}^{i}(y) \rbrace ^{*} d\lambda_{i}
+\lbrace f(x),\phi_{5}^{ai}(y) \rbrace ^{*} d\sigma_{ai}
\\
&+&\lbrace f(x),\phi_{6}^{ai}(y) \rbrace ^{*} d\zeta_{ai}
\Big], 
\end{eqnarray}
where we can observe that the noninvolutive Hamiltonians have been removed. In fact, this is an advantage of the HJ formalism with respect to the Dirac formulation. From one hand, by introducing the generalized brackets  in HJ framework we remove constraints from the beginning. On the other hand, in Dirac's  formulation  we must to  identify  future  constraints by means of consistency, at the end    we need to perform the classification of the constraints in first class and  second class, then Dirac's brackets are introduced and second class constraints can be eliminated;   in HJ scheme we will have at the end   less number of constraints than Dirac's scheme.  \\  
Furthermore, the Frobenius integrability conditions on the  involutive Hamiltonians  $\phi_{3}^{i}$, $\phi_{4}^{i}$, $\phi_{5}^{ai}$ and $\phi_{6}^{ai}$  could introduce new $HJ$ Hamiltonians. In fact, integrability conditions are relevant because ensure that the system is integrable. From the integrability conditions the following Hamiltonians arise 
\begin{eqnarray}
\nonumber
d\phi_{3}^{i}(x)&=&\int d^{3}y\Big[\lbrace \phi_{3}^{i}(x),H'(y)\rbrace ^{*} dt
+\lbrace \phi_{3}^{i}(x),\phi_{3}^{j}(y) \rbrace ^{*} d\tau_{j}
+\lbrace \phi_{3}^{i}(x),\phi_{4}^{j}(y) \rbrace ^{*} d\lambda_{j}
+\lbrace \phi_{3}^{i}(x),\phi_{5}^{aj}(y) \rbrace ^{*} d\sigma_{aj}
\\
&+&\lbrace \phi_{3}^{i}(x),\phi_{6}^{aj}(y) \rbrace ^{*} d\zeta_{aj}
\Big]=0,
\end{eqnarray}
\begin{equation}
\Rightarrow
-(\partial_{a}\pi_{a}{}^{i}-\epsilon^{i}{}_{jk}\pi^{aj}\Upsilon_{a}{}^{k}-\epsilon^{i}{}_{jk}p^{a0j}A_{a0}{}^{k})=0,
\end{equation}
\begin{eqnarray}
\nonumber
d\phi_{4}^{i}(x)&=&\int d^{3}y\Big[\lbrace \phi_{4}^{i}(x),H'(y)\rbrace ^{*} dt
+\lbrace \phi_{4}^{i}(x),\phi_{3}^{j}(y) \rbrace ^{*} d\tau_{j}
+\lbrace \phi_{4}^{i}(x),\phi_{4}^{j}(y) \rbrace ^{*} d\lambda_{j}
+\lbrace \phi_{4}^{i}(x),\phi_{5}^{aj}(y) \rbrace ^{*} d\sigma_{aj}
\\
&+&\lbrace \phi_{4}^{i}(x),\phi_{6}^{aj}(y) \rbrace ^{*} d\zeta_{aj}
\Big]=0,
\end{eqnarray}
\begin{equation}
\Rightarrow
-(\partial_{a}p^{a0i}-\epsilon^{i}{}_{jk}A_{a0}{}^{j}\pi^{ak}-\epsilon^{i}{}_{jk}p^{aoj}\Upsilon_{a}{}^{k})=0,
\end{equation}
\begin{eqnarray}
\nonumber
d\phi_{5}^{ai}(x)&=&\int d^{3}y\Big[\lbrace \phi_{5}^{ai}(x),H'(y)\rbrace ^{*} dt
+\lbrace \phi_{5}^{ai}(x),\phi_{3}^{j}(y) \rbrace ^{*} d\tau_{j}
+\lbrace \phi_{5}^{ai}(x),\phi_{4}^{j}(y) \rbrace ^{*} d\lambda_{j}
+\lbrace \phi_{5}^{ai}(x),\phi_{5}^{bj}(y) \rbrace ^{*} d\sigma_{bj}
\\
&+&\lbrace \phi_{5}^{ai}(x),\phi_{6}^{bj}(y) \rbrace ^{*} d\zeta_{bj}
\Big]=0,
\end{eqnarray}
\begin{equation}
\Rightarrow
\frac{\Xi}{2}\eta^{abc}\left(\partial_{b}A_{c0}{}^{i}-\partial_{c}A_{b0}{}^{i}
+\epsilon^{i}{}_{jk}A_{b0}{}^{j}\Upsilon_{c}{}^{k}-\epsilon^{i}{}_{jk}A_{c0}{}^{j}\Upsilon_{b}{}^{k}\right)
+\frac{1}{2}p^{a0i}=0,
\end{equation}
\begin{eqnarray}
\nonumber
d\phi_{6}^{ai}(x)&=&\int d^{3}y\Big[\lbrace \phi_{5}^{ai}(x),H'(y)\rbrace ^{*} dt
+\lbrace \phi_{6}^{ai}(x),\phi_{3}^{j}(y) \rbrace ^{*} d\tau_{j}
+\lbrace \phi_{6}^{ai}(x),\phi_{4}^{j}(y) \rbrace ^{*} d\lambda_{j}
+\lbrace \phi_{6}^{ai}(x),\phi_{5}^{bj}(y) \rbrace ^{*} d\sigma_{bj}
\\
&+&\lbrace \phi_{6}^{ai}(x),\phi_{6}^{bj}(y) \rbrace ^{*} d\zeta_{bj}
\Big]=0,
\end{eqnarray}
\begin{equation}
\Rightarrow
\frac{\Xi}{2}\eta^{abc}\left(\partial_{b}\Upsilon_{c}{}^{i}-\partial_{c}\Upsilon_{b}{}^{i}
-\epsilon^{i}{}_{jk}A_{b0}{}^{j}A_{c0}{}^{k}+\epsilon^{i}{}_{jk}\Upsilon_{b}{}^{j}\Upsilon_{c}{}^{k}\right)
-\frac{1}{2}\pi^{ai}=0.
\end{equation}
Hence,  we identify the following set of new HJ Hamiltonians
\begin{eqnarray}
\nonumber
\phi_{9}^{i}&\equiv&
\partial_{a}\pi_{a}{}^{i}-\epsilon^{i}{}_{jk}\pi^{aj}\Upsilon_{a}{}^{k}-\epsilon^{i}{}_{jk}p^{a0j}A_{a0}{}^{k}=0,
\\
\nonumber
\phi_{10}^{i}&\equiv&
\partial_{a}p^{a0i}+\epsilon^{i}{}_{jk}\pi^{aj}A_{a0}{}^{k}-\epsilon^{i}{}_{jk}p^{a0j}\Upsilon_{a}{}^{k}=0,
\\
\nonumber
\phi_{11}^{ai}&\equiv&
\frac{\Xi}{2}\eta^{abc}\left(\partial_{b}A_{c0}{}^{i}-\partial_{c}A_{b0}{}^{i}
+\epsilon^{i}{}_{jk}A_{b0}{}^{j}\Upsilon_{c}{}^{k}-\epsilon^{i}{}_{jk}A_{c0}{}^{j}\Upsilon_{b}{}^{k}\right)
+\frac{1}{2}p^{a0i}=0,
\\
\phi_{12}^{ai}&\equiv&
\frac{\Xi}{2}\eta^{abc}\left(\partial_{b}\Upsilon_{c}{}^{i}-\partial_{c}\Upsilon_{b}{}^{i}
-\epsilon^{i}{}_{jk}A_{b0}{}^{j}A_{c0}{}^{k}+\epsilon^{i}{}_{jk}\Upsilon_{b}{}^{j}\Upsilon_{c}{}^{k}\right)
-\frac{1}{2}\pi^{ai}=0. 
\end{eqnarray}
The generalized brackets between these new Hamiltonians are given by 
\begin{eqnarray}
\nonumber
\lbrace \phi_{9}^{i}(x),\phi_{9}^{j}(y) \rbrace^{*}&=&\epsilon^{ij}{}_{k}\phi_{9}^{k}\delta^{3}(x-y),
\\
\nonumber
\lbrace \phi_{9}^{i}(x),\phi_{10}^{j}(y) \rbrace^{*}&=&\epsilon^{ij}{}_{k}\phi_{10}^{k}\delta^{3}(x-y),
\\
\nonumber
\lbrace \phi_{9}^{i}(x),\phi_{11}^{aj}(y) \rbrace^{*}&=&\epsilon^{ij}{}_{k}\phi_{11}^{ak}\delta^{3}(x-y),
\\
\nonumber
\lbrace \phi_{9}^{i}(x),\phi_{12}^{aj}(y) \rbrace^{*}&=&\epsilon^{ij}{}_{k}\phi_{12}^{ak}\delta^{3}(x-y),
\\
\nonumber
\lbrace \phi_{10}^{i}(x),\phi_{10}^{j}(y) \rbrace^{*}&=&-\epsilon^{ij}{}_{k}\phi_{9}^{k}\delta^{3}(x-y),
\\
\nonumber
\lbrace \phi_{10}^{i}(x),\phi_{11}^{aj}(y) \rbrace^{*}&=&\epsilon^{ij}{}_{k}\phi_{12}^{ak}\delta^{3}(x-y),
\\
\nonumber
\lbrace \phi_{10}^{i}(x),\phi_{12}^{bj}(y) \rbrace^{*}&=&-\epsilon^{ij}{}_{k}\phi_{11}^{ak}\delta^{3}(x-y),
\\
\nonumber
\lbrace \phi_{11}^{ai}(x),\phi_{11}^{bj}(y) \rbrace^{*}&=&0,
\\
\nonumber
\lbrace \phi_{11}^{ai}(x),\phi_{12}^{bj}(y) \rbrace^{*}&=&0,
\\
\lbrace \phi_{12}^{ai}(x),\phi_{12}^{bj}(y) \rbrace^{*}&=&0,
\end{eqnarray}
since  the algebra is closed  we conclude that   these  Hamiltonians are involutive, therefore we do not expect new  Hamiltonians. Moreover, we observe that  the Hamiltonians $\phi_{9}^{i}(x)$ and $\phi_{10}^{j}(x)$   are identified  as generators of rotations and  boost respectively. The rest of the Hamiltonians $\phi_{11}(x)$ and $\phi_{12}^{ai}(x)$   are reducible Hamiltonians,  namely, they are not independent;, but a linear  combination of involutive Hamiltonians,  as it will be showed.
With all involutive Hamiltonians and by using the  generalized brackets   the following generalized differential is constructed 
\begin{eqnarray}
\nonumber
df(x)&=&\int d^{3}y\Big[\lbrace f(x),H'(y)\rbrace ^{*} dt
+\lbrace f(x),\phi_{3}^{i}(y) \rbrace ^{*} d\tau_{i}
+\lbrace f(x),\phi_{4}^{i}(y) \rbrace ^{*} d\lambda_{i}
+\lbrace f(x),\phi_{5}^{ai}(y) \rbrace ^{*} d\sigma_{ai}
\\
\nonumber
&+&\lbrace f(x),\phi_{6}^{ai}(y) \rbrace ^{*} d\zeta_{ai}
+\lbrace f(x),\phi_{9}^{i}(y) \rbrace ^{*} d\tilde{\tau}_{i}
+\lbrace f(x),\phi_{10}^{i}(y) \rbrace ^{*} d\tilde{\lambda}_{i}
+\lbrace f(x),\phi_{11}^{ai}(y) \rbrace ^{*} d\tilde{\sigma}_{ai}
\\
&+&\lbrace f(x),\phi_{12}^{ai}(y) \rbrace ^{*} d\tilde{\zeta}_{ai}
\Big],
\label{25a}
\end{eqnarray}
where $\tilde{\tau}_{i}, \tilde{\lambda}_{i}, \tilde{\lambda}_{i}, \tilde{\zeta}_{ai}$ are parameters related with the Hamiltonians  $\phi_{9}^{i}, \phi_{10}^{i}, \phi_{11}^{ai}, \phi_{12}^{ai}$  respectively. Therefore,      from the fundamental differential  we can obtain the relevant symmetries of the theory. In fact, the symmetries are exposed by    the characteristic equations \cite{ F17, F18, F19, F20}, and they are given   by 
\begin{eqnarray}
\nonumber
dA_{a0}{}^{i}&=&\Big[
\epsilon^{ij}{}_{k}A_{a0}{}^{k}T_{j}-\partial_{a}\Lambda^{i}+\epsilon^{ij}{}_{k}\Upsilon_{a}{}^{k}\Lambda_{j}-B_{0a}{}^{0i}
\Big]
dt
\\
&+&\left[\epsilon^{ij}{}_{k}A_{a0}{}^{k}\right]d\tilde{\tau}_{j}
- \left[\delta^{ij}\partial_{a}-\epsilon^{ij}{}_{k}\Upsilon_{a}{}^{k}\right]d\tilde{\lambda}_{j}
+\frac{1}{2}d\tilde{\sigma}_{a}{}^{i},
\end{eqnarray}
\begin{eqnarray}
\nonumber
d\Upsilon_{a}{}^{i}&=&\Big[-
\partial_{a}T^{i}-\epsilon^{ij}{}_{k}\Upsilon_{a}{}^{k}T_{j}-\epsilon^{ij}{}_{k}A_{a0}{}^{k}\Lambda_{j}+B_{0a}{}^{i}\Big]dt
\\
&-&\left[\delta^{ij}\partial_{a}+\epsilon^{ij}{}_{k}\Upsilon_{a}{}^{k}\right]d\tilde{\tau}_{j}
- \left[\epsilon^{ij}{}_{k}A_{a0}{}^{k}\right]d\tilde{\lambda}_{j}
- \frac{1}{2}d\tilde{\zeta}_{a}{}^{i},
\end{eqnarray}
\begin{eqnarray}
dT_{i}&=&d\tau_{i}, \nonumber \\
d\Lambda_{i}&=&d\lambda_{i}, \nonumber \\ 
d\varsigma_{ai}&=&d\sigma_{ai}, \nonumber \\
d\chi_{ai}&=&d\zeta_{ai},
\label{29a}
\end{eqnarray}
\begin{eqnarray}
\nonumber
dB_{ab0}{}^{i}&=&\Big[
\epsilon^{ij}{}_{k}(B_{ab}{}^{k}\Lambda_{j}-B_{ab}{}^{0k}T_{j})+\epsilon^{ij}{}_{k}(\Upsilon_{a}{}^{k}B_{0b}{}^{0}{}_{j}-\Upsilon_{b}{}^{k}B_{0a}{}^{0}{}_{j})-\epsilon^{ij}{}_{k}(A_{a0}{}^{k}B_{0bj}
\\
\nonumber
&-&A_{b0}{}^{k}B_{0aj})-(\partial_{a}B_{b0}{}^{0}{}_{i}-\partial_{b}B_{a0}{}^{0}{}_{i})\Big]dt
\\
\nonumber
&-&\Big[\epsilon^{ij}{}_{k}B_{ab}{}^{k}\Big]d\tilde{\tau}_{j}
\\
\nonumber
&+&\Big[\epsilon^{ij}{}_{k}B_{ab}{}^{0k}\Big]d\tilde{\lambda}_{j}
\\
\nonumber
&+&\frac{1}{2}\left[\delta^{ij}\partial_{a}-\epsilon^{ij}{}_{k}\Upsilon_{a}{}^{k}\right]d\tilde{\sigma}_{bj}-\frac{1}{2}\left[\delta^{ij}\partial_{b}-\epsilon^{ij}{}_{k}\Upsilon_{b}{}^{k}\right]d\tilde{\sigma}_{aj}
\\
&+&\frac{1}{2}\left[\epsilon^{ij}{}_{k}A_{a0}{}^{k}\right]d\tilde{\zeta}_{bj}-\frac{1}{2}\left[\epsilon^{ij}{}_{k}A_{b0}{}^{k}\right]d\tilde{\zeta}_{aj}
\end{eqnarray}
\begin{eqnarray}
\nonumber
dB_{ab}{}^{i}&=&\Big[
-\epsilon^{ij}{}_{k}(B_{ab}{}^{i}T_{j}-B_{ab}{}^{0i}\Lambda_{j})-\epsilon^{ij}{}_{k}(A_{a0}{}^{k}B_{0b}{}^{0}{}_{j}-A_{b0}{}^{k}B_{0a}{}^{0}{}_{j})-\epsilon^{ij}{}_{k}(\Upsilon_{a}{}^{k}B_{0bj}
\\
\nonumber
&-&\Upsilon_{b}{}^{k}B_{0aj})+(\partial_{a}B_{0bi}-\partial_{b}B_{0ai})\Big]dt
\\
\nonumber
&-&\Big[\epsilon^{ij}{}_{k}B_{ab}{}^{k}\Big]d\tilde{\tau}_{j}
\\
\nonumber
&+&\Big[\epsilon^{ij}{}_{k}B_{ab}{}^{0k}\Big]d\tilde{\lambda}_{j}
\\
\nonumber
&+&\frac{1}{2}\left[\epsilon^{ij}{}_{k}A_{c0}{}^{k}\right]d\tilde{\sigma}_{bj}-\frac{1}{2}\left[\epsilon^{ij}{}_{k}A_{b0}{}^{k}\right]d\tilde{\sigma}_{aj}
\\
&-&\frac{1}{2}\left[\delta^{ij}\partial_{a}-\epsilon^{ij}{}_{k}\Upsilon_{a}{}^{k}\right]d\tilde{\zeta}_{bj}+\frac{1}{2}\left[\delta^{ij}\partial_{b}-\epsilon^{ij}{}_{k}\Upsilon_{b}{}^{k}\right]d\tilde{\zeta}_{aj},
\end{eqnarray}
\begin{eqnarray}
d\hat{T}^{i}&=&-\phi_{9}^{i}dt=0, \nonumber \\
d\hat{\Lambda}^{i}&=&-\phi_{10}^{i}dt=0,
\nonumber \\
d\hat{\varsigma}^{i}&=&-\phi_{11}^{ai}dt=0,
\nonumber \\
d\hat{\chi}^{i}&=&-\phi_{12}^{ai}dt=0,
\nonumber \\
dp^{ab0i}&=&0,
\nonumber \\
dp^{abi}&=&0. 
\label{32a}
\end{eqnarray}
From the characteristic equations the following equations of motion arise 
\begin{eqnarray}
\dot{A}_{a0}{}^{i}&=&
{\epsilon}^{ij}{}_{k}A_{a0}{}^{k}T_{j}-\partial_{a}\Lambda^{i}+\epsilon^{ij}{}_{k}\Upsilon_{a}{}^{k}\Lambda_{j}-B_{0a}{}^{0i}, 
\nonumber \\
\dot{\Upsilon}_{a}{}^{i}&=&-
\partial_{a}T^{i}-\epsilon^{ij}{}_{k}\Upsilon_{a}{}^{k}T_{j}-\epsilon^{ij}{}_{k}A_{a0}{}^{k}\Lambda_{j}+B_{0a}{}^{i}, 
\end{eqnarray}
and by taking $dt=0$, the following gauge transformations are identified 
\begin{eqnarray}
\delta A_{a0}{}^{i} &=&\left[\epsilon^{ij}{}_{k}A_{a0}{}^{k}\right]\delta \tilde{\tau}_{j}- \left[\delta^{ij}\partial_{a}-\epsilon^{ij}{}_{k}\Upsilon_{a}{}^{k}\right]\delta \tilde{\lambda}_{j} + \frac{1}{2}\delta \tilde{\sigma}_{a}{}^{i},  \nonumber \\
\delta \Upsilon_{a}{}^{i}&=&- \left[\delta^{ij}\partial_{a}+\epsilon^{ij}{}_{k}\Upsilon_{a}{}^{k}\right] \delta \tilde{\tau}_{j} - \left[\epsilon^{ij}{}_{k}A_{a0}{}^{k}\right]\delta \tilde{\lambda}_{j} - \frac{1}{2}\delta \tilde{\zeta}_{a}{}^{i}.
\end{eqnarray}
Furthermore, we  can identify  from (\ref{29a}) and (\ref{32a}) the  non dynamical variables. In fact,  the former  implies that  the variables $T_{i}, \Lambda_{i}, \varsigma_{ai}, \chi_{ai}$ are   identified as Lagrange multipliers and the later says that the variables $B_{ab0}{}^{i}$ and $B_{ab}{}^{i}$ are non dynamical because the characteristic equations of their  momenta do not present  any evolution. Therefore,  the dynamical variables are given finally by $A_{a0}{}^{i}$ and $\Upsilon_{a}{}^{i}$ and they will be taken into account in the counting of physical degrees of freedom;  all these results were not reported in \cite{7a, 13} \\
On the other hand, we commented above that the involutive constraints $ \phi_{11}^{ai}, \phi_{12}^{ai}$ are reducible; in fact, we can observe that there are 6 reducibility conditions \cite{13}
\begin{eqnarray}
\nonumber
\partial_{a}\phi_{11}^{ai}&=&-\epsilon^{i}{}_{jk}\Upsilon_{a}{}^{j}\phi_{11}^{ak}-\epsilon^{i}{}_{jk}A_{a0}{}^{j}\phi_{12}^{ak}+\frac{1}{2}\phi_{10}^{i},
\\
\partial_{a}\phi_{12}^{ai}&=&\epsilon^{i}{}_{jk}A_{a0}{}^{j}\phi_{11}^{ak}-\epsilon^{i}{}_{jk}\Upsilon_{a}{}^{j}\phi_{12}^{ak}+\frac{1}{2}\phi_{9}^{i}.
\label{33a}
\end{eqnarray}
they are linear combination of involutive Hamiltonians, in this manner, eq.  (\ref{33a})  implies that there are a total of 18  independent involutive Hamiltonians   $\phi_{9}^{i}, \phi_{10}^{i}, \phi_{11}^{ai}, \phi_{12}^{ai}$, and we identified above that there are 18 dynamical variables;  $A_{a0}{}^{i}$ and $\Upsilon_{a}{}^{i}$. Hence, the counting of physical degrees of freedom is carried out as follows; DOF= Dynamical variables - involutive constraints=18-18=0, therefore the theory is devoid of physical degrees of freedom as expected. We can observe in \cite{7a} that in Dirac's approach were used more constraints and dynamical variables in order to  perform the counting of physical degrees  of freedom than in HJ scheme. On the other hand, the HJ generalized brackets and the Fadeev-Jackiw brackets  reported in \cite{7a} coincide to each other, however, in Fadeev-Jakiw scheme we had to fix the gauge in order to obtain the brackets;  in HJ formalism  this was not a necessary step,      in this sense we have completed the results reported in \cite{7a, 13} and we can say  that the HJ scheme is more direct and economical.      
\section{ Hamilton-Jacobi analysis for the  Euler class}
It is well-known that the Euler class can be expressed as a $BF$-like theory \cite{6}
\begin{equation}
S[A_{\mu\nu}{}^{IJ},B_{\mu\nu}{}^{IJ}]=\Omega\int_{M}[*F^{IJ}\wedge B_{IJ}-\frac{1}{2}
*B^{IJ}\wedge B_{IJ}],
\label{34a}
\end{equation}
where $*=\epsilon^{IJKL}$ is the dual of $SO(3,1)$ and $\Omega$ is a constant. The equations of motion arising  from the variation of  (\ref{34a})  are
\begin{eqnarray}
D*B&=&0, \nonumber\\
*F&=&*B, 
\label{35}
\end{eqnarray}
by applying the star product to the equations (\ref{35}) we obtain again (\ref{3}); the Euler and Pontryaging classes  share the same equations of motion. What about their   symmetries?.    By performing the  3+1 decomposition, break down  the Lorentz covariance and introducing the variables defined in previous sections  we obtain that the Euler  Lagrangian  takes the form 
\begin{eqnarray}
\nonumber
\mathcal{L}&=&\Omega\eta^{abc}\int_{M}\Big[
B_{ab}{}^{0i}\dot{\Upsilon}_{ci}+B_{abi}\dot{A}_{c}{}^{0i}
\\
\nonumber
&-&\left(\partial_{a}B_{bc}{}^{0i}-\epsilon^{i}{}_{jk}B_{bc}{}^{0j}\Upsilon_{a}{}^{k}+\epsilon^{i}{}_{jk}A_{a}{}^{0j}B_{bc}{}^{k}\right)T_{i}
\\
\nonumber
&+&\left(\partial_{a}B_{bc}{}^{i}+\epsilon^{i}{}_{jk}\Upsilon_{a}{}^{j}B_{bc}{}^{k}+\epsilon^{i}{}_{jk}A_{a0}{}^{j}B_{bc}{}^{0k}\right)\Lambda_{i}
\\
\nonumber
&-&\frac{1}{2}\left(\partial_{a}\Upsilon_{b}{}^{i}-\partial_{b}\Upsilon_{a}{}^{i}-\epsilon^{i}{}_{jk}A_{a0}{}^{j}A_{b0}{}^{k}+\epsilon^{i}{}_{jk}\Upsilon_{a}{}^{j}\Upsilon_{b}{}^{k}-B_{ab}{}^{i}\right)\varsigma_{ci}
\\
&+&\frac{1}{2}\left(\partial_{a}A_{b0}{}^{i}-\partial_{b}A_{a0}{}^{i}+\epsilon^{i}{}_{jk}A_{a0}{}^{j}\Upsilon_{b}{}^{k}-\epsilon^{i}{}_{jk}A_{b0}{}^{j}\Upsilon_{a}{}^{k}+B_{ab}{}^{0i}\right)\chi_{ci}\Big]d^{3}x,
\end{eqnarray}
now, in order to compare  the symmetries of both theories, we will  use   the same phase space variables  that we used  in the Pontryagin invariant. Hence, by using  the following Hamiltonians for Euler class
\begin{eqnarray}
\nonumber
H'&\equiv&\Pi+H_{0}=0
\\
\nonumber
\phi_{1}^{a0i}&\equiv& p^{a0i}-\Xi\eta^{abc}B_{bc}{}^{0i}=0,
\\
\nonumber
\phi_{2}^{ai}&\equiv& \pi^{ai}-\Xi\eta^{abc}B_{bc}{}^{i}=0,
\\
\nonumber
\phi_{3}^{i}&\equiv&\hat{T}^{i}=0,
\\
\nonumber
\phi_{4}^{i}&\equiv&\hat{\Lambda}^{i}=0,
\\
\nonumber
\phi_{5}^{ai}&\equiv&\hat{\varsigma}^{ai}=0,
\\
\nonumber
\phi_{6}^{ai}&\equiv&\hat{\chi}^{ai}=0,
\\
\nonumber
\phi_{7}^{ab0i}&\equiv& p^{ab0i}=0,
\\
\phi_{8}^{abi}&\equiv& p^{abi}=0,
\label{37}
\end{eqnarray}
where the canonical Hamiltonian for  the  Euler  invariant  given in terms of the Hamiltonians (\ref{37})  is expressed as 
\begin{eqnarray}
\nonumber
H_{0}&=&\frac{\Omega}{\Xi}\Big[
\partial_{a}p^{a0i}+\epsilon^{i}{}_{jk}\pi^{aj}A_{a0}{}^{k}-\epsilon^{i}{}_{jk}p^{a0j}\Upsilon_{a}{}^{k}\Big]T_{i}
-\frac{\Omega}{\Xi}\Big[\partial_{a}\pi^{ai}-\epsilon^{i}{}_{jk}\pi^{aj}\Upsilon_{a}{}^{k}-\epsilon^{i}{}_{jk}p^{a0j}A_{a0}{}^{k}\Big]\Lambda_{i}
\\
\nonumber
&+&\frac{\Omega}{2}\eta^{abc}\Big[\partial_{b}\Upsilon_{c}{}^{i}-\partial_{c}\Upsilon_{b}{}^{i}-\epsilon^{i}{}_{jk}A_{b0}{}^{j}A_{c0}{}^{k}+\epsilon^{i}{}_{jk}\Upsilon_{b}{}^{j}\Upsilon_{c}{}^{k}\Big]\varsigma_{ai}
\\
\nonumber
&-&\frac{\Omega}{2}\eta^{abc}\Big[\partial_{b}A_{c0}{}^{i}-\partial_{c}A_{b0}{}^{i}+\epsilon^{i}{}_{jk}A_{b0}{}^{j}\Upsilon_{c}{}^{k}-\epsilon^{i}{}_{jk}A_{c0}{}^{j}\Upsilon_{b}{}^{k}\Big]\chi_{ai}
\\
&-&\frac{1}{2}\frac{\Omega}{\Xi}\varsigma_{ai}\pi^{ai}-\frac{1}{2}\frac{\Omega}{\Xi}\chi_{ai}p^{a0i}, 
\end{eqnarray}
then, from the Lagrangian  we identify the following fundamental Poisson brackets
\begin{eqnarray}
\nonumber
\lbrace A_{a0i}(x),\pi^{bj}(y)\rbrace &=&-\frac{\Xi}{\Omega}\delta_{a}^{b}\delta_{i}^{j}\delta^{3}(x-y),
\\
\nonumber
\lbrace \Upsilon_{ai}(x),p^{b0j}(y)\rbrace &=&\frac{\Xi}{\Omega}\delta_{a}^{b}\delta_{i}^{j}\delta^{3}(x-y),
\\
\nonumber
\lbrace T_{i}(x),\hat{T}^{j}(y)\rbrace &=&\delta_{a}^{b}\delta_{i}^{j}\delta^{3}(x-y),
\\
\nonumber
\lbrace \Lambda_{i}(x),\hat{\Lambda}^{j}(y)\rbrace &=&\delta_{a}^{b}\delta_{i}^{j}\delta^{3}(x-y),
\\
\nonumber
\lbrace \varsigma_{ai}(x),\hat{\varsigma}^{bj}(y)\rbrace &=&\delta_{a}^{b}\delta_{i}^{j}\delta^{3}(x-y),
\\
\nonumber
\lbrace \chi_{ai}(x),\hat{\chi}^{bj}(y)\rbrace &=&\delta_{a}^{b}\delta_{i}^{j}\delta^{3}(x-y),
\\
\nonumber
\lbrace B_{ab0i}(x),p^{cd0j}(y)\rbrace&=&\frac{1}{2}(\delta_{a}^{c}\delta_{b}^{d}-\delta_{a}^{d}\delta_{b}^{c})\delta_{i}^{j}\delta^{3}(x-y),
\\
\lbrace B_{abi}(x),p^{cdj}(y)\rbrace &=&\frac{1}{2}(\delta_{a}^{c}\delta_{b}^{d}-\delta_{a}^{d}\delta_{b}^{c})\delta_{i}^{j}\delta^{3}(x-y),
\end{eqnarray}
 it is appreciable in these brackets the contribution of the constants $\Omega$ and  $\Xi$. Moreover, the role of the canonically conjugate   variables has changed; now  $\pi^{bj}$ is canonical conjugated  to $  A_{a0i}$ and $p^{b0j}$ is canonical conjugated   to $\Upsilon_{ai}$  whereas  in Pontryagin they were interchanged.   Furthermore, by using the HJ Hamiltonians  (\ref{37}) the fundamental differential for the Euler theory is given by 
\begin{eqnarray}
\nonumber
df(x)&=&\int d^{3}y\Big[\lbrace f(x),H'(y)\rbrace dt
+\lbrace f(x),\phi_{1}^{a0i} \rbrace d\omega_{a0i}
+\lbrace f(x),\phi_{2}^{ai} \rbrace d\varphi_{ai}
+\lbrace f(x),\phi_{3}^{i} \rbrace d\tau_{i}
+\lbrace f(x),\phi_{4}^{i} \rbrace d\lambda_{i}
\\
&+&\lbrace f(x),\phi_{5}^{ai} \rbrace d\sigma_{ai}
+\lbrace f(x),\phi_{6}^{ai} \rbrace d\zeta_{ai}
+\lbrace f(x),\phi_{7}^{ab0i} \rbrace d\theta_{ab0i}
+\lbrace f(x),\phi_{8}^{abi} \rbrace d\xi_{abi}
\Big],
\end{eqnarray}
where $\omega_{a0i}, \varphi_{ai},  \tau_{i},  \lambda_{i}, \sigma_{ai}, \zeta_{ai}, \theta_{ab0i}, \xi_{abi}$ are parameters related with the Hamiltonians. Moreover, we can notice that $\phi_{3}^{i}$, $\phi_{4}^{i}$, $\phi_{5}^{ai}$ and $\phi_{6}^{ai}$ are involutives and $\phi_{1}^{a0i}$, $\phi_{2}^{ai}$ $\phi_{7}^{ab0i}$ and $\phi_{8}^{abi}$ are noninvolutives. In this manner, noninvolutives Hamiltonians allow us to introduce the generalized brackets (\ref{10}), where 
\begin{equation}
C_{\bar{a}\bar{b}}
=
\begin{pmatrix}
0 & 0 & \Xi\eta^{abc}\delta^{il} & 0 \\
0 & 0 & 0 & -\Xi\eta^{abc}\delta^{il} \\
-\Xi\eta^{abc}\delta^{il} & 0 & 0 & 0 \\
0 & \Xi\eta^{abc}\delta^{il} & 0 & 0 \\
\end{pmatrix}
\delta^{3}(x-y),
\end{equation}

with inverse given by

\begin{equation}
C_{\bar{a}\bar{b}}^{-1}
=
\begin{pmatrix}
0 & 0 & -\frac{1}{2\Xi}\eta^{def}\delta_{lj} & 0 \\
0 & 0 & 0 & \frac{1}{2\Xi}\eta^{def}\delta_{lj} \\
\frac{1}{2\Xi}\eta_{def}\delta_{lj} & 0 & 0 & 0 \\
0 & -\frac{1}{2\Xi}\eta_{def}\delta_{lj} & 0 & 0 \\
\end{pmatrix}
\delta^{3}(x-y),
\end{equation}
hence, the generalized brackets are given by 
\begin{eqnarray}
\nonumber
\lbrace A_{a0i}(x),\pi^{bj}(y)\rbrace ^{*} &=&-\frac{\Xi}{\Omega}\delta_{a}^{b}\delta_{i}^{j}\delta^{3}(x-y),
\\
\nonumber
\lbrace \Upsilon_{ai}(x),p^{b0j}(y)\rbrace ^{*} &=&\frac{\Xi}{\Omega}\delta_{a}^{b}\delta_{i}^{j}\delta^{3}(x-y),
\\
\nonumber
\lbrace T_{i}(x),\hat{T}^{j}(y)\rbrace ^{*} &=&\delta_{a}^{b}\delta_{i}^{j}\delta^{3}(x-y),
\\
\nonumber
\lbrace \Lambda_{i}(x),\hat{\Lambda}^{j}(y)\rbrace ^{*} &=&\delta_{a}^{b}\delta_{i}^{j}\delta^{3}(x-y),
\\
\nonumber
\lbrace \varsigma_{ai}(x),\hat{\varsigma}^{bj}(y)\rbrace ^{*} &=&\delta_{a}^{b}\delta_{i}^{j}\delta^{3}(x-y),
\\
\nonumber
\lbrace \chi_{ai}(x),\hat{\chi}^{bj}(y)\rbrace ^{*} &=&\delta_{a}^{b}\delta_{i}^{j}\delta^{3}(x-y),
\\
\nonumber
\lbrace B_{ab0i}(x),p^{cd0j}(y)\rbrace ^{*} &=&0,
\\
\nonumber
\lbrace B_{abi}(x),p^{cdj}(y)\rbrace ^{*} &=&0,
\\
\nonumber
\lbrace B_{ab0i}(x),\Upsilon_{cj}(y)\rbrace ^{*} &=&\frac{1}{2\Omega}\eta_{abc}\delta_{ij}\delta^{3}(x-y),
\\
\lbrace B_{abi}(x),A_{c0j}(y)\rbrace ^{*} &=&\frac{1}{2\Omega}\eta_{abc}\delta_{ij}\delta^{3}(x-y),
\label{44a}
\end{eqnarray}
we can observe that  the Euler and Pontryagin theories  have different generalized brackets;  the  contribution of the constants $\Omega$ and $\Xi$ is manifested. Furthermore,  the generalized brackets (\ref{44a}) coincide with those reported in \cite{13}  where   the Fadeev-Jackiw approach was developed, in addition,  for obtaining those brackets  a temporal  gauge fixing was necessary.   The introduction of the generalized brackets allow us to introduce a new fundamental differential
\begin{eqnarray}
\nonumber
df(x)&=&\int d^{3}y\Big[\lbrace f(x),H'(y)\rbrace ^{*} dt
+\lbrace f(x),\phi_{3}^{i}(y) \rbrace ^{*} d\tau_{i}
+\lbrace f(x),\phi_{4}^{i}(y) \rbrace ^{*} d\lambda_{i}
+\lbrace f(x),\phi_{5}^{ai}(y) \rbrace ^{*} d\sigma_{ai}
\\
&+&\lbrace f(x),\phi_{6}^{ai}(y) \rbrace ^{*} d\zeta_{ai}
\Big].
\end{eqnarray}
On the other hand, integrability conditions on the involutive Hamiltonians $\phi_{3}^{i}$, $\phi_{4}^{i}$, $\phi_{5}^{ai}$ and $\phi_{6}^{ai}$ introduce new $HJ$ Hamiltonians. In fact, from integrability  conditions the following Hamiltonians arise 
\begin{eqnarray}
\nonumber
d\phi_{3}^{i}(x)&=&0
\quad
\Rightarrow 
\:
\phi_{9}^{i}: \frac{\Omega}{\Xi}(\partial_{a}p^{a0i}+\epsilon^{i}{}_{jk}\pi^{aj}A_{a0k}-\epsilon^{i}{}_{jk}p^{a0j}\Upsilon_{a}{}^{k})=0,
\\
\nonumber
d\phi_{4}^{i}(x)&=&0
\quad
\Rightarrow
\:
\phi_{10}^{i}:\frac{\Omega}{\Xi}(\partial_{a}\pi^{ai}-\epsilon^{i}{}_{jk}\pi^{aj}\Upsilon_{a}{}^{k}-\epsilon^{i}{}_{jk}p^{a0j}A_{a0}{}^{k})=0,
\\
\nonumber
d\phi_{5}^{ai}(x)&=&0
\quad
\Rightarrow
\:
\phi_{11}^{ai}:-\frac{\Omega}{2}\eta^{abc}(\partial_{b}\Upsilon_{c}{}^{i}-\partial_{c}\Upsilon_{b}{}^{i}-\epsilon^{i}{}_{jk}A_{b0}{}^{j}A_{c0}{}^{k}+\epsilon^{i}{}_{jk}\Upsilon_{b}{}^{j}\Upsilon_{c}{}^{k})+\frac{1}{2}\frac{\Omega}{\Xi}\pi^{ai}=0,
\\
d\phi_{6}^{ai}(x)&=&0
\quad
\Rightarrow
\:
\phi_{12}^{ai}: \frac{\Omega}{2}\eta^{abc}(\partial_{b}A_{c0}{}^{i}-\partial_{c}A_{b0}{}^{i}+\epsilon^{i}{}_{jk}A_{b0}{}^{j}\Upsilon_{c}{}^{k}-\epsilon^{i}{}_{jk}A_{c0}{}^{j}\Upsilon_{b}{}^{k})+\frac{1}{2}\frac{\Omega}{\Xi}p^{a0i}=0, \nonumber \\
\label{45}
\end{eqnarray}
the generalized algebra between these new Hamiltonians read 
\begin{eqnarray}
\nonumber
\lbrace \phi_{9}^{i}(x),\phi_{9}^{j}(y) \rbrace^{*}&=&\epsilon^{ij}{}_{k}\phi_{9}^{k}\delta^{3}(x-y),
\\
\nonumber
\lbrace \phi_{9}^{i}(x),\phi_{10}^{j}(y) \rbrace^{*}&=&\epsilon^{ij}{}_{k}\phi_{10}^{k}\delta^{3}(x-y),
\\
\nonumber
\lbrace \phi_{9}^{i}(x),\phi_{11}^{aj}(y) \rbrace^{*}&=&\epsilon^{ij}{}_{k}\phi_{11}^{k}\delta^{3}(x-y),
\\
\nonumber
\lbrace \phi_{9}^{i}(x),\phi_{12}^{aj}(y) \rbrace^{*}&=&\epsilon^{ij}{}_{k}\phi_{12}^{k}\delta^{3}(x-y),
\\
\nonumber
\lbrace \phi_{10}^{i}(x),\phi_{10}^{j}(y) \rbrace^{*}&=&-\epsilon^{ij}{}_{k}\phi_{9}{}^{k}\delta^{3}(x-y),
\\
\nonumber
\lbrace \phi_{10}^{i}(x),\phi_{11}^{aj}(y) \rbrace^{*}&=&\epsilon^{ij}{}_{k}\phi_{12}^{ak}\delta^{3}(x-y),
\\
\nonumber
\lbrace \phi_{10}^{i}(x),\phi_{12}^{bj}(y) \rbrace^{*}&=&-\epsilon^{ij}{}_{k}\phi_{11}^{ak}\delta^{3}(x-y),
\\
\nonumber
\lbrace \phi_{11}^{ai}(x),\phi_{11}^{bj}(y) \rbrace^{*}&=&0,
\\
\nonumber
\lbrace \phi_{11}^{ai}(x),\phi_{12}^{bj}(y) \rbrace^{*}&=&0,
\\
\lbrace \phi_{12}^{ai}(x),\phi_{12}^{bj}(y) \rbrace^{*}&=&0, 
\end{eqnarray}
which  is closed,   therefore we do not expect new  Hamiltonians. Furthermore, now we observe that  the Hamiltonians  $\phi_{9}^{i}(x)$ and   $\phi_{10}^{j}(x)$   are identified  as generators of rotations and  boost respectively.  This is a relevant result  because the generators for Euler theory  are interchanged with respect  Pontryagin's invariant; this result  implies that the theories are classically different. The Hamiltonians (\ref{45}) are involutives and this fact allows  us to introduce a new generalized differential 
\begin{eqnarray}
\nonumber
df(x)&=&\int d^{3}y\Big[\lbrace f(x),H'(y)\rbrace ^{*} dt
+\lbrace f(x),\phi_{3}^{i}(y) \rbrace ^{*} d\tau_{i}
+\lbrace f(x),\phi_{4}^{i}(y) \rbrace ^{*} d\lambda_{i}
+\lbrace f(x),\phi_{5}^{ai}(y) \rbrace ^{*} d\sigma_{ai}
\\
\nonumber
&+&\lbrace f(x),\phi_{6}^{ai}(y) \rbrace ^{*} d\zeta_{ai}
+\lbrace f(x),\phi_{9}^{i}(y) \rbrace ^{*} d\tilde{\tau}_{i}
+\lbrace f(x),\phi_{10}^{i}(y) \rbrace ^{*} d\tilde{\lambda}_{i}
+\lbrace f(x),\phi_{11}^{ai}(y) \rbrace ^{*} d\tilde{\sigma}_{ai}
\\
&+&\lbrace f(x),\phi_{12}^{ai}(y) \rbrace ^{*} d\tilde{\zeta}_{ai}
\Big],
\label{48a}
\end{eqnarray}
where $\tilde{\tau}_{i}, \tilde{\lambda}_{i}, \tilde{\sigma}_{ai}, \tilde{\zeta}_{ai} $ are parameters associated to the Hamiltonians (\ref{45}). It is worth to comment   that the  generalized brackets (\ref{44a}) make the fundamental differentials (\ref{25a}) and   (\ref{48a})  to be completely different. In fact, due to the generalized brackets of the theories are different to each other,  the fundamental differentials will describe  different scenarios  on the phase space.     From the generalized differential (\ref{48a})   we can obtain the characteristic equations of the theory,  then we can  identify the symmetries. The characteristic equations are given by 
\begin{eqnarray}
dA_{a0}{}^{i}&=&\Big[
\epsilon^{ij}{}_{k}A_{a0}{}^{k}T_{j}-\partial_{a}\Lambda^{i}+\epsilon^{ij}{}_{k}\Upsilon_{a}{}^{k}\Lambda_{j}+\frac{1}{2}\varsigma_{a}{}^{i}
\Big]
dt
\nonumber \\
&+& \left[\epsilon^{ij}{}_{k}A_{a0}{}^{k}\right]d\tilde{\tau}_{j}
+\left[\delta^{ij}\partial_{a}-\epsilon^{ij}{}_{k}\Upsilon_{a}{}^{k}\right]d\tilde{\lambda}_{j}
+ \frac{1}{2}d\tilde{\sigma}_{a}{}^{i},
\label{48}
\end{eqnarray}

\begin{eqnarray}
\nonumber
d\Upsilon_{a}{}^{i}&=&\Big[
-\partial_{a}T^{i}+\epsilon^{ij}{}_{k}\Upsilon_{a}{}^{k}T_{j}-\epsilon^{ij}{}_{k}A_{a0}{}^{k}\Lambda_{j}-\frac{1}{2}\chi_{a}{}^{i}\Big]dt
\\
&-&\left[\delta^{ij}\partial_{a}-\epsilon^{ij}{}_{k}\Upsilon_{a}{}^{k}\right]d\tilde{\tau}_{j}
- \left[\epsilon^{ij}{}_{k}A_{a0}{}^{k}\right]d\tilde{\lambda}_{j}
+\frac{1}{2}d\tilde{\zeta}_{a}{}^{i},
\label{49}
\end{eqnarray}
\begin{eqnarray}
dT_{i}&=&d\tau_{i}, \nonumber \\
d\Lambda_{i}&=&d\lambda_{i},
 \nonumber \\
d\varsigma_{ai}&=&d\sigma_{ai},
\nonumber \\
d\chi_{ai}&=&d\zeta_{ai},
\label{51}
\end{eqnarray}
\begin{eqnarray}
\nonumber
dB_{ab0}{}^{i}&=&\Big[
\epsilon^{ij}{}_{k}(B_{ab}{}^{k}\Lambda_{j}-B_{ab}{}^{0k}T_{j})+\epsilon^{ij}{}_{k}(\Upsilon_{a}{}^{k}B_{0b}{}^{0}{}_{j}-\Upsilon_{b}{}^{k}B_{0a}{}^{0}{}_{j})-\epsilon^{ij}{}_{k}(A_{a0}{}^{k}B_{0bj}
\\
\nonumber
&-&A_{b0}{}^{k}B_{0aj})-(\partial_{a}B_{0b}{}^{0}{}_{j}-\partial_{b}B_{0a}{}^{0}{}_{j})\Big]dt
\\
\nonumber
&-&\Big[\epsilon^{ij}{}_{k}B_{ab}{}^{0k}\Big]d\tilde{\tau}_{j}
\\
\nonumber
&+&\Big[\epsilon^{ij}{}_{k}B_{ab}{}^{k}\Big]d\tilde{\lambda}_{j}
\\
\nonumber
&+&\frac{1}{2}\left[\delta^{ij}\partial_{a}-\epsilon^{ij}{}_{k}\Upsilon_{a}{}^{k}\right]d\tilde{\sigma}_{bj}-\frac{1}{2}\left[\delta^{ij}\partial_{b}-\epsilon^{ij}{}_{k}\Upsilon_{b}{}^{k}\right]d\tilde{\sigma}_{aj}
\\
&+&\frac{1}{2}\left[\epsilon^{ij}{}_{k}A_{a0}{}^{k}\right]d\tilde{\zeta}_{bj}-\frac{1}{2}\left[\epsilon^{ij}{}_{k}A_{b0}{}^{k}\right]d\tilde{\zeta}_{aj}
\\
\nonumber
\end{eqnarray}

\begin{eqnarray}
\nonumber
dB_{ab}{}^{i}&=&\Big[
\epsilon^{ij}{}_{k}(B_{ab}{}^{k}T_{j}+B_{ab}{}^{0k}\Lambda_{j})-\epsilon^{ij}{}_{k}(A_{a0}{}^{k}B_{0a}{}^{0}{}_{j}-A_{b0}{}^{k}B_{0b}{}^{0}{}_{j})-\epsilon^{ij}{}_{k}(\Upsilon_{a}{}^{k}B_{0bj}
\\
\nonumber
&-&\Upsilon_{b}{}^{k}B_{0aj})+(\partial_{a}B_{0bj}-\partial_{b}B_{0aj})\Big]dt
\\
\nonumber
&+&\Big[\epsilon^{ij}{}_{k}B_{ab}{}^{k}\Big]d\tilde{\tau}_{j}
\\
\nonumber
&+&\Big[\epsilon^{ij}{}_{k}B_{ab}{}^{0k}\Big]d\tilde{\lambda}_{j}
\\
\nonumber
&+&\frac{1}{2}\left[\epsilon^{ij}{}_{k}A_{a0}{}^{k}\right]d\tilde{\sigma}_{bj}-\frac{1}{2}\left[\epsilon^{ij}{}_{k}A_{b0}{}^{k}\right]d\tilde{\sigma}_{aj}
\\
&+&\frac{1}{2}\left[\delta^{ij}\partial_{a}-\epsilon^{ij}{}_{k}\Upsilon_{a}{}^{k}\right]d\tilde{\zeta}_{bj}-\frac{1}{2}\left[\delta^{ij}\partial_{b}-\epsilon^{ij}{}_{k}\Upsilon_{b}{}^{k}\right]d\tilde{\zeta}_{aj},
\end{eqnarray}
\begin{eqnarray}
d\hat{T}^{i}&=&-\phi_{9}^{i}dt=0,
\nonumber \\
d\hat{\Lambda}^{i}&=&-\phi_{10}^{i}dt=0,
\nonumber \\
d\hat{\varsigma}^{i}&=&-\phi_{11}^{ai}dt=0,
\nonumber \\
d\hat{\chi}^{i}&=&+\phi_{12}^{ai}dt=0,
\nonumber \\
dp^{ab0i}&=&0,
\nonumber \\
dp^{abi}&=&0, 
\label{54}
\end{eqnarray}
thus, from the characteristics equations we identify the Euler's equations of motions given by 
\begin{eqnarray}
\dot{A}_{a0}{}^{i}&=& \epsilon^{ij}{}_{k}A_{a0}{}^{k}T_{j}-\partial_{a}\Lambda^{i}+\epsilon^{ij}{}_{k}\Upsilon_{a}{}^{k}\Lambda_{j}+\frac{1}{2}\varsigma_{a}{}^{i}, \nonumber \\
\dot{\Upsilon}_{a}{}^{i}&=& -\partial_{a}T^{i}+\epsilon^{ij}{}_{k}\Upsilon_{a}{}^{k}T_{j}-\epsilon^{ij}{}_{k}A_{a0}{}^{k}\Lambda_{j}-\frac{1}{2}\chi_{a}{}^{i}, 
\end{eqnarray}
and by taking $dt=0$ the following gauge transformations arise
\begin{eqnarray}
\delta A_{a0}{}^{i}&=& \left[\epsilon^{ij}{}_{k}A_{a0}{}^{k}\right]\delta \tilde{\tau}_{j}  + \left[\delta^{ij}\partial_{a}-\epsilon^{ij}{}_{k}\Upsilon_{a}{}^{k}\right]\delta \tilde{\lambda}_{j} + \frac{1}{2}\delta \tilde{\sigma}_{a}{}^{i}, \nonumber \\
\delta \Upsilon_{a}{}^{i}&=&- \left[\delta^{ij}\partial_{a}-\epsilon^{ij}{}_{k}\Upsilon_{a}{}^{k}\right]\delta \tilde{\tau}_{j}
- \left[\epsilon^{ij}{}_{k}A_{a0}{}^{k}\right]\delta \tilde{\lambda}_{j} + \frac{1}{2}\delta \tilde{\zeta}_{a}{}^{i}.
\end{eqnarray}
Moreover, from eq. (\ref{51})  we can observe that  $T_{i}, \Lambda_{i}, \varsigma_{ai}, \chi_{ai}$ are   identified as Lagrange multipliers and (\ref{54})   says  that the variables $B_{ab0}{}^{i}$ and $B_{ab}{}^{i}$ are not dynamical. Therefore,  the dynamical variables are given by $A_{a0}{}^{i}$ and $\Upsilon_{a}{}^{i}$ just like in Pontryaging invariant. \\
On the other hand, we observe that the constrains $ \phi_{11}^{ai}, \phi_{12}^{ai}$ are not independent and present the following 6 reducibility conditions
\begin{eqnarray}
\nonumber
\partial_{a}\phi_{11}^{ai}&=&-\epsilon^{i}{}_{jk}\Upsilon_{a}{}^{j}\phi_{11}^{ak}+\epsilon^{i}{}_{jk}A_{a0}{}^{j}\phi_{12}^{ak}-\frac{1}{2}\phi_{10}^{i},
\\
\partial_{a}\phi_{12}^{ai}&=&-\epsilon^{i}{}_{jk}A_{a0}{}^{j}\phi_{11}^{ak}-\epsilon^{i}{}_{jk}\Upsilon_{a}{}^{j}\phi_{12}^{ak}+\frac{1}{2}\phi_{9}^{i}.
\label{53}
\end{eqnarray}
hence,  eq. (\ref{48}) and  eq. (\ref{49})  implies that there are a total of 18  dynamical variables, and there are 18 independent involutive constraints  $\phi_{9}^{i}, \phi_{10}^{i}, \phi_{11}^{ai}, \phi_{12}^{ai}$. It is worth to mention  that the reducibility  conditions (\ref{33a}) and (\ref{53})  do not affect integrability. In  fact, these conditions are linear  combination of involutive Hamiltonians, then the generalized brackets between reducibility conditions  with  other Hamiltonians vanish. Thus, the counting of physical degrees of freedom  yield  to conclude that  the Euler theory lacks of physical degrees of freedom.   
\section{Conclussions}
In this paper a complete  HJ analysis for the Pontryagin  and Euler classes has been performed.  We have developed  our study by using in both theories the same phase space variables and  this fact has allowed us to compare the emergent  symmetries.  The full set of Hamiltonians of the theories  were identified and different  generalized HJ differentials have  been constructed. From the generalized  differential all symmetries of the theories have been found, we observed that in spite of the invariants share the same equations of motion and the same dynamical variables, the symmetries  are different. In fact, the generators  of rotations and boost are interchanged; the generators of  boost  and rotations   for Pontryagin are generators of rotations and boost for  Euler respectively. Moreover, we found that the generalized brackets are also different because there is a direct  contribution of the parameters $\Omega$ and $\Xi$, which could be relevant  in the quantization process or the  identification   of the observables of the theories. In this respect, the generalized brackets between the dynamical variables for both theories   are not the same to each other, and this implies that the corresponding  uncertainty principles  will be different. It is worth to comment,  that all these results will be important when a boundary is  added. In fact, we can see in \cite{8} that the knowledge  of the canonical structure of topological theories with a boundary  is  a mandatory step to perform in order to know the symmetries and physical degrees of freedom at the boundary, thus,   due to the close relation between Pontryagin and Euler  invariants  a  work  is in  progress \cite{19}.   \\
In this manner,  we can  observe that our results are generic and we have extended those reported in \cite {7a, 13};     we have also  showed that the HJ formulation is an  elegant and  complete framework for studying topological gauge theories.   \\
\noindent \textbf{Acknowledgements}\\[1ex]
We would like to thank R. Cartas-Fuentevilla for discussions and reading of the manuscript. 

\end{document}